\documentclass[twocolumn,superscriptaddress,nofootinbib,floatfix,aps,showpacs,prl,citeautoscript,reprint]{revtex4-1}
\usepackage{dcolumn}
\usepackage{bm}
\usepackage{amsmath}
\usepackage{amsfonts}
\usepackage{amssymb}
\usepackage{graphicx}
\usepackage{tabularx}
\usepackage{subfigure}
\usepackage{color}
\usepackage{gensymb}
\usepackage{physics}
\usepackage{multirow}

\usepackage{hyperref} \hypersetup{colorlinks=true, linkcolor=blue,
  citecolor=blue, urlcolor=blue, pdftitle={Low energy phases of bilayer Bi predicted by structure search in two dimensions}}

\newcommand{\minus}{\scalebox{0.75}[1.0]{$-$}}

\newcolumntype{c}[1]{>{\centering\arraybackslash}p{#1}}

\begin{document}

\title{Low energy phases of bilayer Bi predicted by structure search in two dimensions}

\author{Sobhit Singh}
\email{smsingh@mix.wvu.edu}
\affiliation{Department of Physics and Astronomy, West Virginia University, Morgantown, WV 26505-6315, USA}
\affiliation{Department of Physics and Astronomy, Rutgers University, Piscataway, NJ 08854, USA}

\author{Zeila Zanolli}
\affiliation{Catalan Institute of Nanoscience and Nanotechnology (ICN2) and European Theoretical Spectroscopy Facility, CSIC and BIST, Campus UAB, Bellaterra, 08193 Barcelona, Spain}

\author{Maximilian Amsler}
\affiliation{Laboratory of Atomic and Solid State Physics, Cornell University, Ithaca, New York 14853, USA}

\author{B. Belhadji}
\affiliation{NanoMat/Q-Mat/CESAM and European Theoretical Spectroscopy Facility, Universit{`e} de Li{\`e}ge (B5), B-4000 Li{\`e}ge, Belgium}

\author{Jorge O. Sofo}
\affiliation{The Pennsylvania State University, 201 Old Main, University Park, Pennsylvania 16802, USA}

\author{Matthieu J. Verstraete}
\affiliation{NanoMat/Q-Mat/CESAM and European Theoretical Spectroscopy Facility, Universit{`e} de Li{\`e}ge (B5), B-4000 Li{\`e}ge, Belgium}

\author{Aldo H. Romero}
\email{Aldo.Romero@mail.wvu.edu}
\affiliation{Department of Physics and Astronomy, West Virginia University, Morgantown, WV 26505-6315, USA}

\begin{abstract}
We employ an {\it ab-initio} structure search algorithm to explore the configurational space of Bi in quasi two dimensions. A confinement potential restricts the movement of atoms within a pre-defined thickness during structure search calculations within the minima hopping method to find the stable and metastable forms of bilayer Bi. In addition to recovering the two known low-energy structures (puckered monoclinic and buckled hexagonal), our calculations predict three new structures of bilayer Bi. We call these structures the $\alpha$, $\beta$, and $\gamma$ phases of bilayer Bi, which are, respectively, 63, 72, and 83\,meV/atom higher in energy than that of the monoclinic ground state, and thus potentially synthesizable using appropriate substrates. We also compare the structural, electronic, and vibrational properties of the different phases. The puckered monoclinic, buckled hexagonal, and $\beta$ phases exhibit a semiconducting energy gap, whereas $\alpha$ and $\gamma$ phases are metallic. We notice an unusual Mexican-hat type band dispersion leading to a van Hove singularity in the buckled hexagonal bilayer Bi. Notably, we find symmetry-protected topological Dirac points in the electronic spectrum of the $\gamma$ phase. The new structures suggest that bilayer Bi provides a novel playground to study distortion-mediated metal-insulator phase transitions. 
\end{abstract}


\keywords{Bismuth, bilayer, structure search, two-dimensions, puckered structure, minima hopping method}

\maketitle

\section{Introduction}
Two dimensional (2D) materials made from group V of the periodic table (pnictogens = N, P, As, Sb, Bi) have attracted much attention due to their unique electronic and topological properties~\cite{butler_ACSreview2013, zhenPRL2014, riveroPRB2014, HeineC4CS00102H2014, ciraciNPRB2015, wuphos_PRB2015, Kou_JPCL2015, balendhran2015elemental, sobhitPRB2017}. Experimental synthesis of phosphorene (a single layer of black phosphorus), which was the first addition to the elemental 2D material family after graphene, has further spurred research into group V elemental 2D materials~\cite{HanLiu2014, LikaiLi2014, Churchill2014}. Recent theoretical predictions as well as the experimental studies revealed the existence of stable 2D nitrogene, arsenene, antimonene, and bismuthene~\cite{Liu_ACSNano2014, zhenPRL2014, riveroPRB2014, ciraciNPRB2015, wuphos_PRB2015, Kou_JPCL2015, balendhran2015elemental, Arsenene_PRB2015, Ciraci_PRB2015, Ciraci_PRB2016, Ersan_PRB2016, Udo_PRB2016, Martin2017}. Among the pnictogens, Bi is particularly interesting due to its strong intrinsic spin-orbit coupling (SOC) and thermoelectric properties. Large Rashba effect and non-trivial topological phases have been observed in Bi thin films~\cite{KoroteevPRL2004, HiraharaPRl2006, ChristianPRL2007, KoroteevPRB2008, HJKim2015, YandongMa2015, YunhaoLuNanoLett2015}. Reis {\it et al.} reported the presence of quantum spin Hall effects together with a large energy gap in Bi films grown on SiC(0001)~\cite{Reis287}. Bi also shows enhanced thermoelectricity in reduced dimensions~\cite{Heremans1999, DresselhausMS2007}.

The geometric structure of bulk Bi is based on a trigonal pyramid ($R$-$3m$ space group). Indeed, the outermost shell of Bi atom has $6s^{2}6p^{3}$ electronic configuration and it tends to form three covalent bonds with neighboring atoms to complete its shell. Hence, Bi atomic layers are naturally expected to have a buckled hexagonal structure. However, recent experiments and theoretical studies suggest that below 4 atomic monolayer thickness Bi prefers a puckered monoclinic structure, similar to phosphorene~\cite{NagaoPRL2004, SCOTT2005175, YunhaoLuNanoLett2015, Fangeaaq0330}. Moreover, a single atomic sheet of Bi contains out-of-plane dangling bonds. These dangling bonds mutually saturate in a bilayer stacking. Hence, even numbers of Bi layers are energetically more favorable than the odd ones~\cite{YunhaoLuNanoLett2015}. The puckered monoclinic phase of bilayer Bi has been experimentally synthesized on various substrates~\cite{NagaoPRL2004, SCOTT2005175, YunhaoLuNanoLett2015, Fangeaaq0330}. However, depending upon the growth conditions and choice of substrate, Bi atoms could yield distinct structural arrangements~\cite{HRSharma2008}. Therefore, it is important to understand the energetics of different stable and metastable structures of bilayer Bi. 
Many global optimization methods have been developed to perform structure search calculations for bulk materials~\cite{RandomSearch, Metadynamics-2, simulated_annealing-1, basin_hopping, evolutionary_metadynamics, genetic_algo-1, genetic_algo-2, evolutionary-algorithm, particle-swarm, Guillermo}, and a few have been developed specifically for structural search in two dimensions~\cite{WYanchaoJPC2012, XiaojunACSN2012, Haigang2013, HongjunJACS2014, YinchangJunPRB2016, AlexanderPRB2016, RevardPRB2016, MaximilianCoM2017, ArunimaPRB2017}. Finally, in the past two years, datasets and databases have been deployed for 2D materials~\cite{Marzari2Dnature_nanotech, Thygesen2018, RasmussenThygesen2015, KChoudhary2017, Midwest2Ddatabase, AshtonPRL2017, JTPaulHennig2017}, but contain principally known 2D materials or derivatives of known 3D ones.
 
\begin{figure*}[htb!]
 \centering
 \includegraphics[width=17.5cm, keepaspectratio=true]{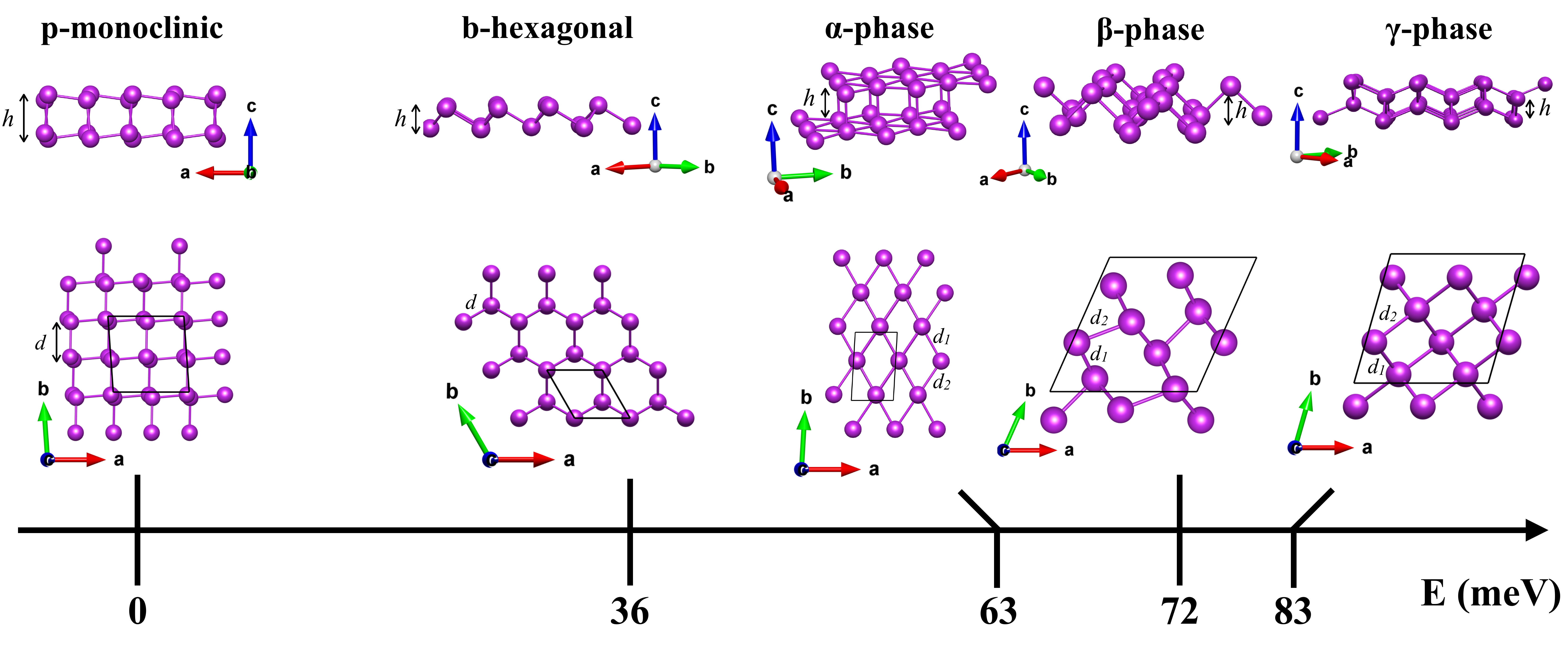}
 \caption{(Color online) Energetic ordering and relaxed structures (side and top views) of the low-energy phases of bilayer Bi obtained from the constrained MHM calculations. The energy differences between the structures (in meV/atoms) are reported with respect to the formation energy of the p-mono structure. }
 \label{fig:crystal}
 \end{figure*}

In the present work, we systematically explore the multidimensional potential energy landscape of bilayer Bi, using a constrained minima hopping method. We predict three new structures ($\alpha$, $\beta$, and $\gamma$ phases), in addition to recovering the two known phases of bilayer Bi: puckered monoclinic (p-mono) and buckled hexagonal (b-hexa). Our calculations predict the puckered monoclinic structure as the ground state of bilayer Bi, which is consistent with previous studies~\cite{NagaoPRL2004, YunhaoLuNanoLett2015, Ciraci_PRB2016}. The b-hexa, $\alpha$, $\beta$, and $\gamma$ phases of bilayer Bi are metastable. We predict that the $\gamma$-phase hosts topologically non-trivial Dirac points in its electronic spectrum. In their free standing form, the p-mono, b-hexa, and $\beta$ phases are dynamically stable, whereas the $\alpha$ and $\gamma$ phases exhibit imaginary phonon frequencies implying dynamical instability. All the predicted phases might be stabilized under the effect of epitaxy on a substrate, and uniaxial or biaxial strain. We further discuss the structural, electronic, and vibrational properties of all the obtained low-energy structures, together with the details of the 2D structure search method we employ.

\begin{figure*}[htb!]
 \centering
 \includegraphics[width=17.5cm, keepaspectratio=true]{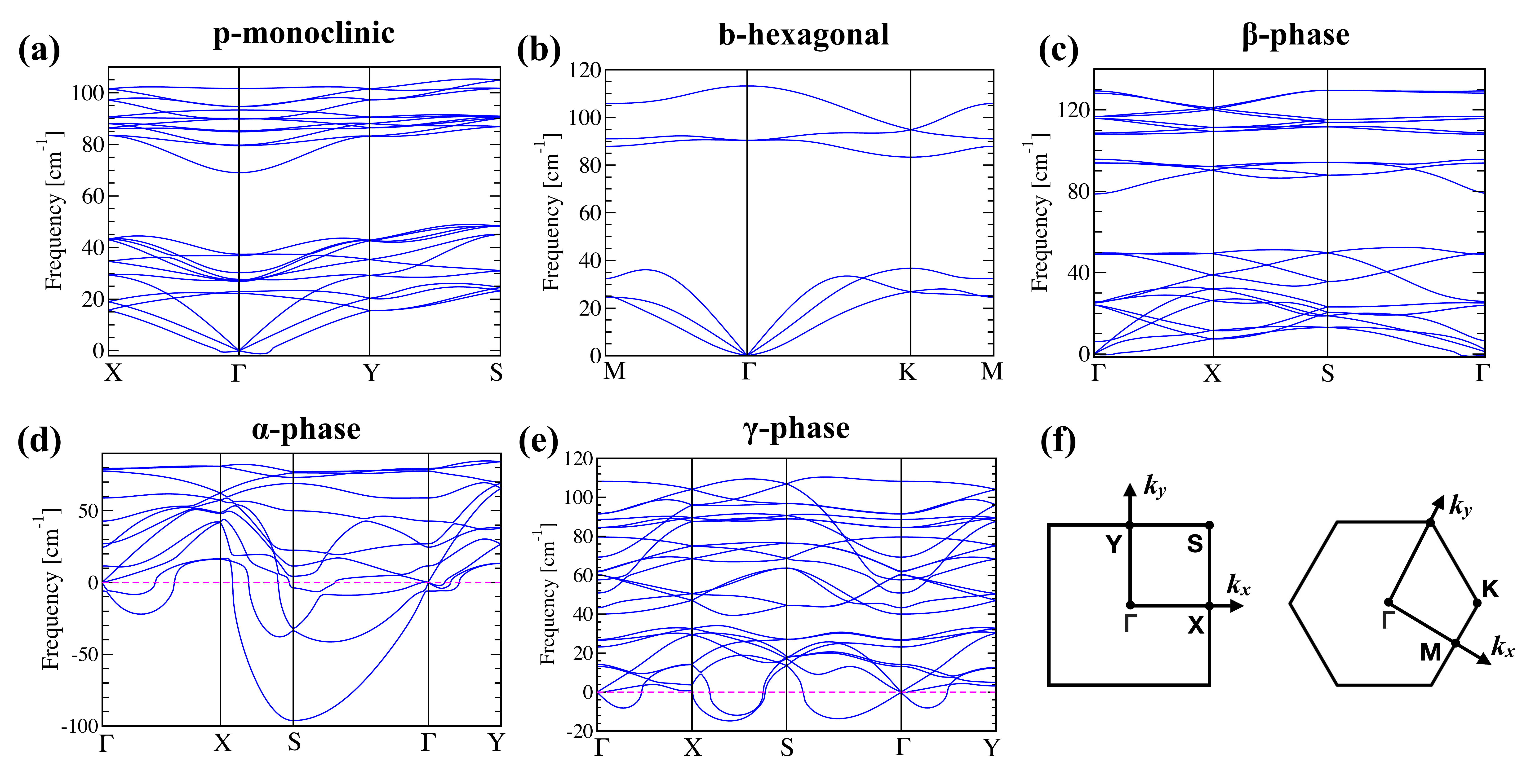}
 \caption{(Color online) Bilayer phonons calculated along the high-symmetry directions in momentum space. 2D Brillouin zones for square and hexagonal lattice are shown in (f). The $\alpha$ phase slightly breaks the acoustic sum rule due to the numerical differences between the MHM and PHONOPY calculations.  }
 \label{fig:phonons}
 \end{figure*}

\begin{figure*}[hb!]
 \centering
 \includegraphics[width=14cm, keepaspectratio=true]{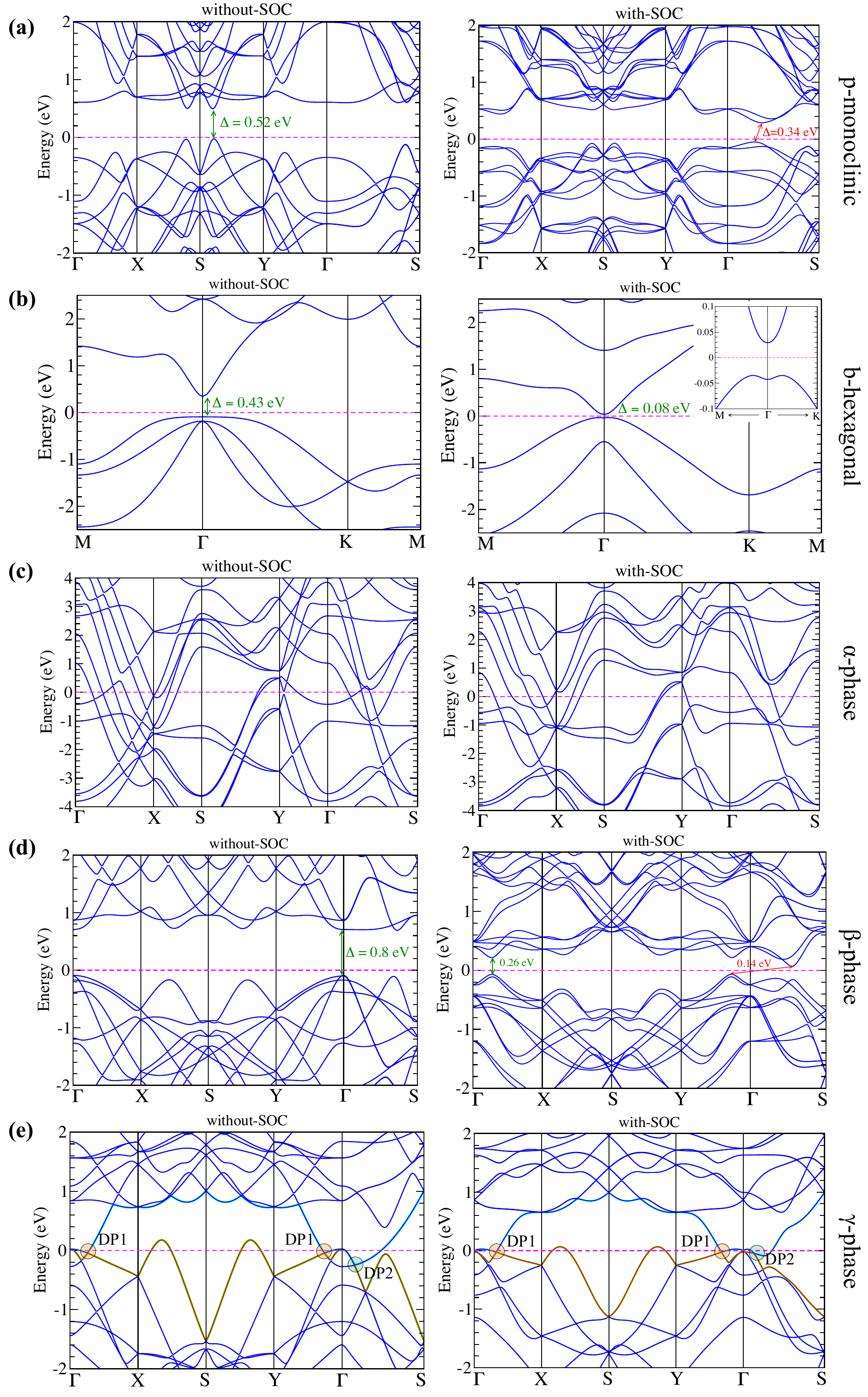}
 \caption{(Color online) Electronic bands calculated with (right panel) and without (left panel) inclusion of spin-orbit coupling for bilayer Bi in (a) p-monoclinic, (b) b-hexagonal, (c) $\alpha$-, (d) $\beta$-, and (e) $\gamma$-phases. Dashed magenta line depicts the Fermi-level and $\Delta$ denotes the direct energy bandgap in semiconducting bilayers. Light colors highlight the spin-degeneracy of bands near the Fermi-level in (e). }
 \label{fig:bands}
 \end{figure*}

\begin{table*}[htb!]
\small
\begin{center}
    \renewcommand*{\arraystretch}{1.25}
\caption{Number of atoms per unit cell ($N_a$), lattice parameters, cell angles, bond lengths, and formation energy of the obtained low-energy structures of bilayer Bi \\ }\label{tab:crys_details}

\begin{tabular}{ m{2.0cm}  m{1.5cm} m{3.9cm} m{3.9cm} m{3.5cm} m{1.5cm} m{2.0cm} }

\hline\\[-3mm]
\multirow{2}{*}{Structure} & $N_a$ &  lattice parameters  & 	cell angles 	& Bi-Bi~bond~length 	& $h$  & E$_{form}$   \\ 
					&	& (in \AA)			& 	(in degrees)	&   (in \AA)	&	(in \AA)	&	(in eV/atom)		\\ 
\hline\\[-3mm]
p-monoclinic & 8 &	$a=$ 6.7224,~$b=$ 6.7224	&	$\alpha=\beta=$ 90,\,$\gamma=$ 94		&	$d$ = 3.11	&	3.10 		&	0.106  \\
\hline\\[-3mm]

b-hexagonal  &	2 & $a=$ 4.5984,~$b=$ 4.5984		&	$\alpha=\beta=$ 90,\,$\gamma=$ 120 	&	$d$ = 3.12  	&	1.63 		&	0.142  \\
\hline\\[-3mm]

$\alpha$-phase &  4 & $a=$ 3.5605,~ $b=$ 5.7778 	&	$\alpha=$ 90.7,\,$\beta=$ 89.4,\,$\gamma=$ 86.7  	&	$d_1$ = 3.30,\,$d_2$ = 3.48 	&	3.16 	& 0.169	 \\
\hline\\[-3mm]

$\beta$-phase 	& 8 & $a=$ 7.9550,~$b=$ 7.9446 	& $\alpha=\beta=$ 90,\,$\gamma=$ 65.9		&	$d_1$ = 3.05,\,$d_2$ = 3.15 	&	2.14 		&	0.178		 \\
\hline\\[-3mm]

$\gamma$-phase & 8  & $a=$ 7.6127,~$b=$ 7.6127		& $\alpha=\beta=$ 90,\,$\gamma=$ 74.1	&	$d_1$ = 3.10,\,$d_2$ = 3.31    	&	2.08 		&	0.190	  \\
\hline

\end{tabular}
\end{center}
\end{table*}

\section{Results and Discussion}
Figure~\ref{fig:crystal} shows the optimized crystal structures and formation energies of the five low-energy configurations of bilayer Bi predicted by the constrained MHM. Our MHM calculations recover the reported p-mono ground state and b-hexa first metastable structure~\cite{NagaoPRL2004, YunhaoLuNanoLett2015, Ciraci_PRB2016}, but also predict three new phases of bilayer Bi which are 63, 72, and 83\,meV/atom higher in energy. Generally, group V elements (N, P, As, Sb, Bi) tend to form a puckered structure in 2D. Although bulk Bi prefers a rhombohedral structure with alternate stacking of hexagonal buckled Bi bilayers, the b-hexa bilayer is energetically less favorable compared to the p-mono structure by 36 meV/atom. 

Crystallographic details and formation energies ($E_{form}$) are summarized in Table~\ref{tab:crys_details}. 
The optimized bilayer structures are also provided in the supplemental material (SM)~\cite{supplemental}. We calculate $E_{form}$ using the formula, $E_{form} = \frac{E_{bilayer}}{n_{bilayer}} \minus \frac{E_{bulk}}{n_{bulk}}$, where $E_{bilayer}$ and $E_{bulk}$ represent the total energy of the bilayer structures and bulk Bi ($R$-$3m$), respectively, and $n_{bilayer}$ ($n_{bulk}$) denotes the number of atoms in the respective unit cells. The formation energy also gives us an estimate of the interlayer strength in the layered structures. For most of the synthesized single-layer transition-metal dichalcogenides (TMDs), $E_{form}$ ranges from 80 -- 150 meV/atom~\cite{HZhuangJPCC2013}, and for the hexagonal group III--V single layer materials $E_{form}$ ranges from 380 -- 520\,meV/atom~\cite{HennigZhuangPRB2013}. As an important benchmark, the $E_{form}$ of single layers of ZnO and silicene are 190 and 760\,meV/atom, and both have been successfully synthesized~\cite{TuschePRL2007, SilicenePRL2012, ZhuangHennig2014}. In our case, the $\gamma$ phase has the highest formation energy (83\,meV/atom), therefore, it should be within the reach of experimental synthesis. 

In order to test the dynamical stability of the different phases, we calculate their phonon frequencies in the 2D Brillouin zone, shown in Fig.~\ref{fig:phonons}. We find that the p-mono, b-hexa, and $\beta$ phases are dynamically stable. The very small dynamical instabilities visible near the $\Gamma$ point are numerical and linked to the 2D out-of-plane acoustic mode ($ZA$) with quadratic dispersion, which is not perfectly reproduced for $k \rightarrow 0$~\cite{HimadriAIP2015, sobhitPRB2017, SobhitPRBG_MoS22018}. The quadratic dispersion of the $ZA$ phonon mode will become linear in $k$ under strain~\cite{HimadriAIP2015}.

The instability of the $\alpha$ and $\gamma$ phases is not unexpected as they are highly symmetric buckled rectangular lattices (see Fig.~\ref{fig:crystal}): Bi has three valence orbitals ($p_x$, $p_y$, $p_z$), and should form three covalent bonds to saturate its valency. This requirement is met in p-mono, b-hexa and $\beta$ phases of bilayer Bi, but not in the $\alpha$ and $\gamma$ phases, where the coordination number is four for each Bi atom. Consequently, the systems will be metallic and a structural instability is highly probable for the $\alpha$ and $\gamma$ phases at ambient conditions. Importantly, there are demonstrations that structures with anomalous coordination numbers can be stabilized under pressure or strain conditions~\cite{Takarabe1988, MCMILLAN1999171, Lapidus2013}.

\begin{figure}[htb!]
 \centering
 \includegraphics[width=9.0cm, keepaspectratio=true]{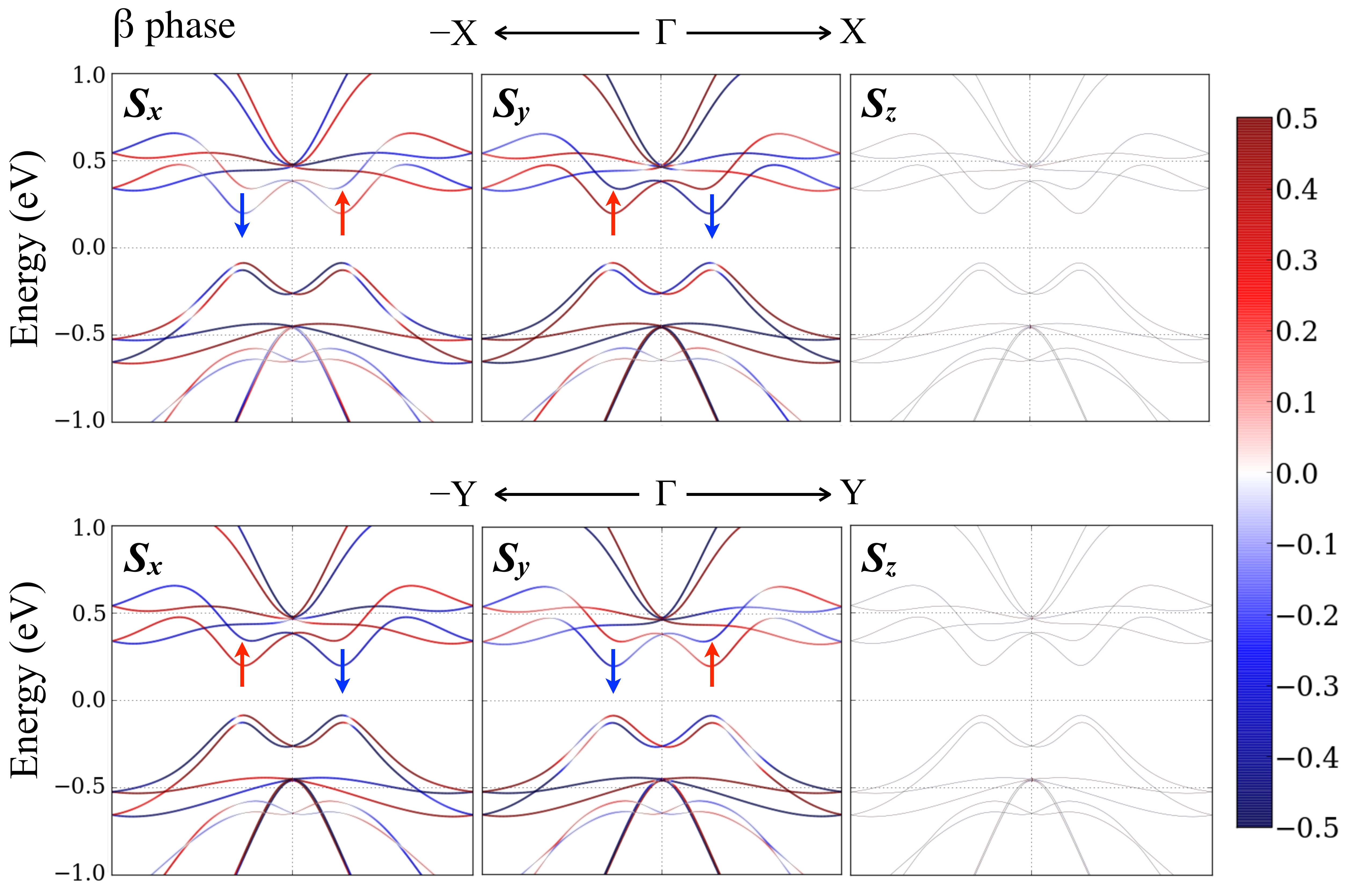}
 \caption{(Color online) Spin projected electronic band structure for $\beta$ phase. Red/Blue color depicts spin up/down states. }
 \label{fig:bands_spinBETA}
 \end{figure}

Figure~\ref{fig:bands} shows the electronic band structure of all five phases along the high-symmetry directions, with and without SOC. The p-mono, b-hexa and $\beta$ phases are semiconducting, whereas $\alpha$ and $\gamma$ phases are metallic. We observe a direct energy bandgap (DFT-PBE) of 0.52, 0.43 and 0.8\,eV in p-mono, b-hexa and $\beta$ phases, respectively. The strong SOC effects of bismuth reduce the bandgap of the semiconducting phases, to 0.34, 0.08, and 0.14\,eV, and change the direct gap nature of p-mono and $\beta$ to indirect (transition marked as a red arrow in Fig.~\ref{fig:bands} (a, d)). The SOC induces a spin-splitting of electronic bands in the non-centrosymmetric p-mono and $\beta$ phases. No such spin-splitting occurs in the b-hexa phase due to the protected inversion-symmetry. However, the top valence band inherits a Mexican-hat type dispersion near the $\Gamma$ point (inset of Fig.~\ref{fig:bands}(b)) which leads to a van Hove singularity in the density of states near the Fermi-level ($E_F$)~\cite{SeixasPRL2016, OzdamarPRB2018}. This is particularly interesting because a small amount of charge doping (hole doping) will trigger time-reversal symmetry breaking and may give rise to emergent phenomena such as ferromagnetism, ferroelasticity, multiferroicity, or superconductivity in two dimensions~\cite{SeixasPRL2016}.

Our electron band structures for b-hexa differ from those reported by Akt\"urk {\it et al.}~\cite{Ciraci_PRB2016} due to our explicit inclusion of semi-core $5d^{10}$ electrons (we also use VASP and PBE+SOC). Their predicted bandgap of 0.547\,eV is much larger than our 0.08\,eV, and their optimized lattice parameters of 4.38\,\AA, is perceptibly smaller than our 4.598\,\AA.
In the SM~\cite{supplemental}, we replicate the results of Ref.~\cite{Ciraci_PRB2016} without including Bi $5d$ electrons ($i.e.$ considering Bi $5d$ electrons in the core). For the p-mono phase, the PBE+SOC predicted bandgap changes only from 0.34\,eV to 0.39\,eV upon freezing the Bi $5d^{10}$ electrons in the core. Overall, the shape of the electronic bands is similar for both PAW datasets (with and without $5d^{10}$ valence electrons), and we believe the calculations with explicit $d$ electrons should produce more accurate results as it has been reported for bulk Bi~\cite{singh2016PCCP,  ssPRBelastic_2018}.

Valley spin-splitting effects are observed in the electronic band structure of the $\beta$-phase along the $\Gamma$-X and $\Gamma$-Y directions. Figure~\ref{fig:bands_spinBETA} shows the spin-projected electronic band structure of the $\beta$-phase calculated with-SOC along $\minus X \rightarrow \Gamma \rightarrow X$ and $\minus Y \rightarrow \Gamma \rightarrow Y$ directions. The spin polarization is entirely contained in the $x-y$ plane, with two valleys polarized along $-x + y$ and the other two polarized along $+x -y$ due to the $C_{2}$  rotational symmetry. The conduction and valence band edges have opposite spin polarization in both directions. The two lowest conduction and two highest valence bands are all composed of $p_{z}$ orbitals. The dispersion and spin-texture of the lowest conduction band and two highest valence bands in Fig.~\ref{fig:bands_spinBETA} resemble the electronic band structure of single layer transition metal dichalcogenides $MX_{2}$ ($M = \{Mo, W\}$, and $X = \{S, Se\}$), in which spin-valley effects have been observed~\cite{WangYaoPRB2008, SallenPRB2012, MakNature2012, XiaoPRL2012, HualingNature2012}. The broken inversion-symmetry and strong SOC effects lift the spin degeneracy of bands everywhere except at the Kramer's points of the $\beta$-phase. The time-reversal symmetry further couples the spin and valley degrees of freedom of valleys located at $\pm {{\bf k}}$, yielding valley-specific optical selection rules.  This is similar to the case of $MX_{2}$ monolayers~\cite{WangYaoPRB2008, SallenPRB2012, XiaoPRL2012, singh2018proximity}, but with in-plane instead of out-of-plane spin texture, which may yield novel spin-pseudospin and magnetic valley couplings. The corresponding optical transitions can be probed in photoluminescence measurements using circularly polarized light. 

As mentioned above the $\alpha$ and $\gamma$ phases are metallic due to the unsaturated $p$-orbitals. A small structural distortion or charge instability could break this coordination of Bi atoms, and might lead to a structural phase transition into one of the lower symmetry phases of bilayer Bi, which are semiconducting. On the other hand, epitaxial deposition on a substrate may also stabilize the different phases.

\begin{figure}[htb!]
 \centering
 \includegraphics[width=8.5cm, keepaspectratio=true]{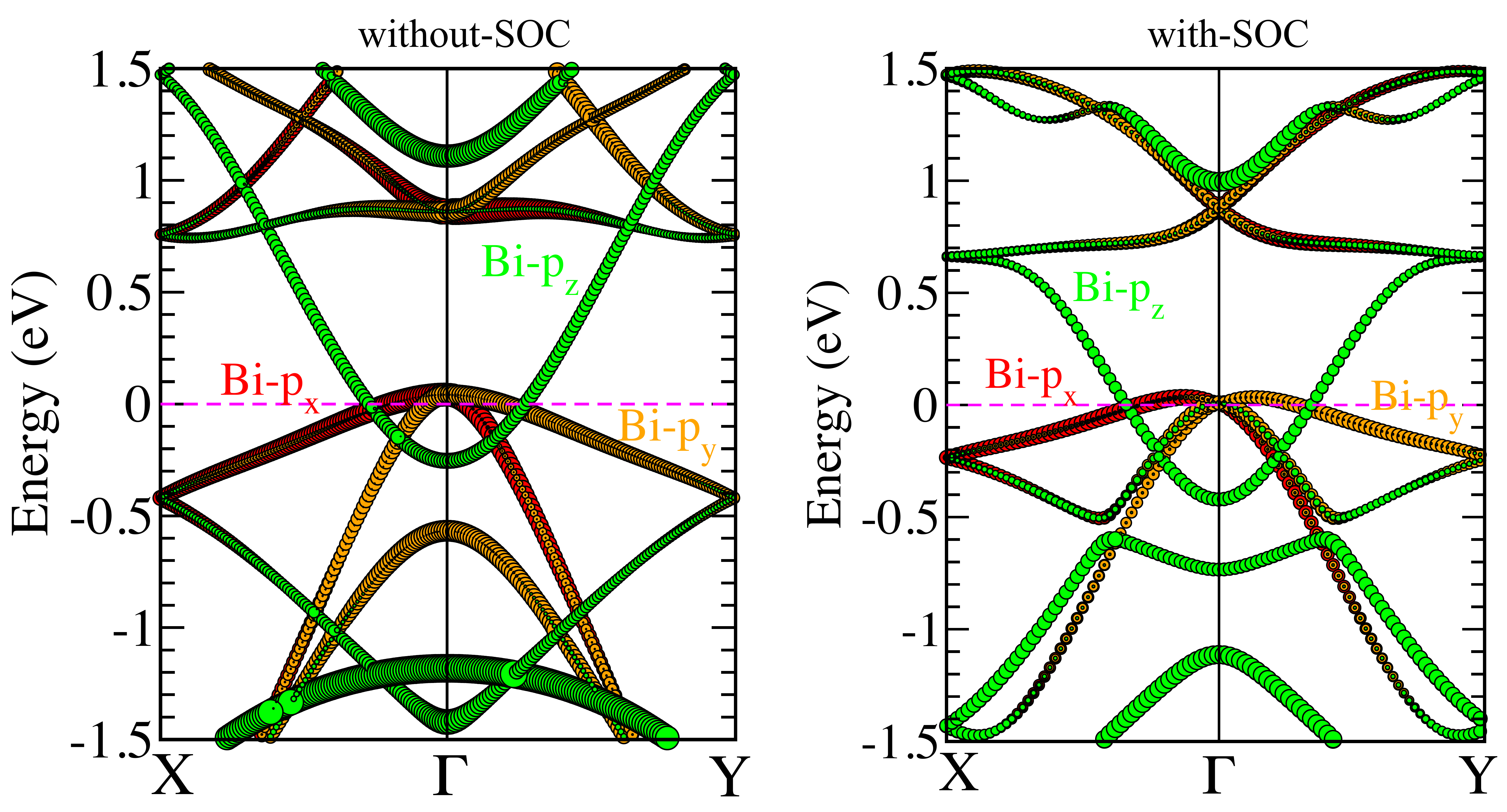}
 \caption{(Color online) Atomic orbital projected electronic band structure of $\gamma$ phase calculated without-SOC (left panel) and with-SOC (right panel).}
 \label{fig:bands_orbitalsDNL}
 \end{figure}

Interestingly, we find that the $\gamma$ phase hosts topologically protected gapless type-I Dirac points (DP1) near the Fermi-level, as shown in Fig.~\ref{fig:bands}(e) along $\Gamma \rightarrow X$ and $\Gamma \rightarrow Y$, at 40 meV below $E_{F}$. The DP1 points result from an inverted band-ordering of Bi-$p_{z}$ (conduction) and Bi-$p_{x,y}$ (valence) bands near the $\Gamma$ point (Fig.~\ref{fig:bands_orbitalsDNL}). Since these bands belong to different 1D irreducible representations, they are allowed to cross along the high-symmetry line. These band-crossings near the Fermi-level are topologically protected by the vertical mirror symmetries of the bilayer. Without SOC, the direct coordinates of these 4-fold degenerate Dirac points in momentum space are: ($\pm$0.13, 0, 0) and (0, $\pm$0.13, 0). These points DP1 could belong to a Dirac nodal line centered at $\Gamma$, but our band structure calculation shown in Fig.~\ref{fig:bands}(e) reveal that there is no such DP1 along the $\Gamma \rightarrow S$. However, we notice a 6-fold degenerate Dirac point (DP2) located along the $\Gamma \rightarrow S$ at a lower energy than that of DP1. The direct coordinates of DP2 in the energy and momentum space (without-SOC) are $E_{F} \minus$ 0.25\,eV and ($\pm$0.08, $\pm$0.08, 0), respectively.

Strong SOC effects of Bi slightly shift the location of the 4-fold degenerate DP1 away from the $\Gamma$ point along the symmetry axis to new coordinates: ($\pm$0.167, 0, 0) and (0, $\pm$0.167, 0). On the contrary, SOC effects shift the DP2 closer to the $\Gamma$ point to new coordinates ($\pm$0.065, $\pm$0.065, 0), and also lift the degeneracy of DP2 from 6-fold to 4-fold. Moreover, in presence of SOC DP1 and DP2 move closer to the Fermi-level and reside at energies $E_{F} \minus$ 0.025\,eV and $E_{F} \minus$ 0.045\,eV, respectively. Analysis of the Fermi-velocity near the band crossing points (with-SOC) suggests a type-I nature of DP1 and type-II nature of tilted DP2~\cite{ShengyuanPRB2017}. Type-II Dirac points usually appear at the  touching points of electron and hole pockets located near the Fermi-level. The formation of DP2 can be clearly seen in Fermi surface plots shown in the SM~\cite{supplemental}. We also observe signatures of interesting topological Lifshitz transitions with changes in the chemical potential near the Fermi-level (see SM~\cite{supplemental}). This suggests the occurrence of distinct topological phase transitions in the $\gamma$ phase~\cite{GEVolovik2017}. The presence of such intriguing changes in the electronic band structure near the Fermi-level and changes in the Fermi-surface topology calls for a dedicated and more rigorous investigation of the topological features present in the $\gamma$ phase, which is beyond the scope of present work. 

\begin{figure}[htb!]
 \centering
 \includegraphics[width=7cm, keepaspectratio=true]{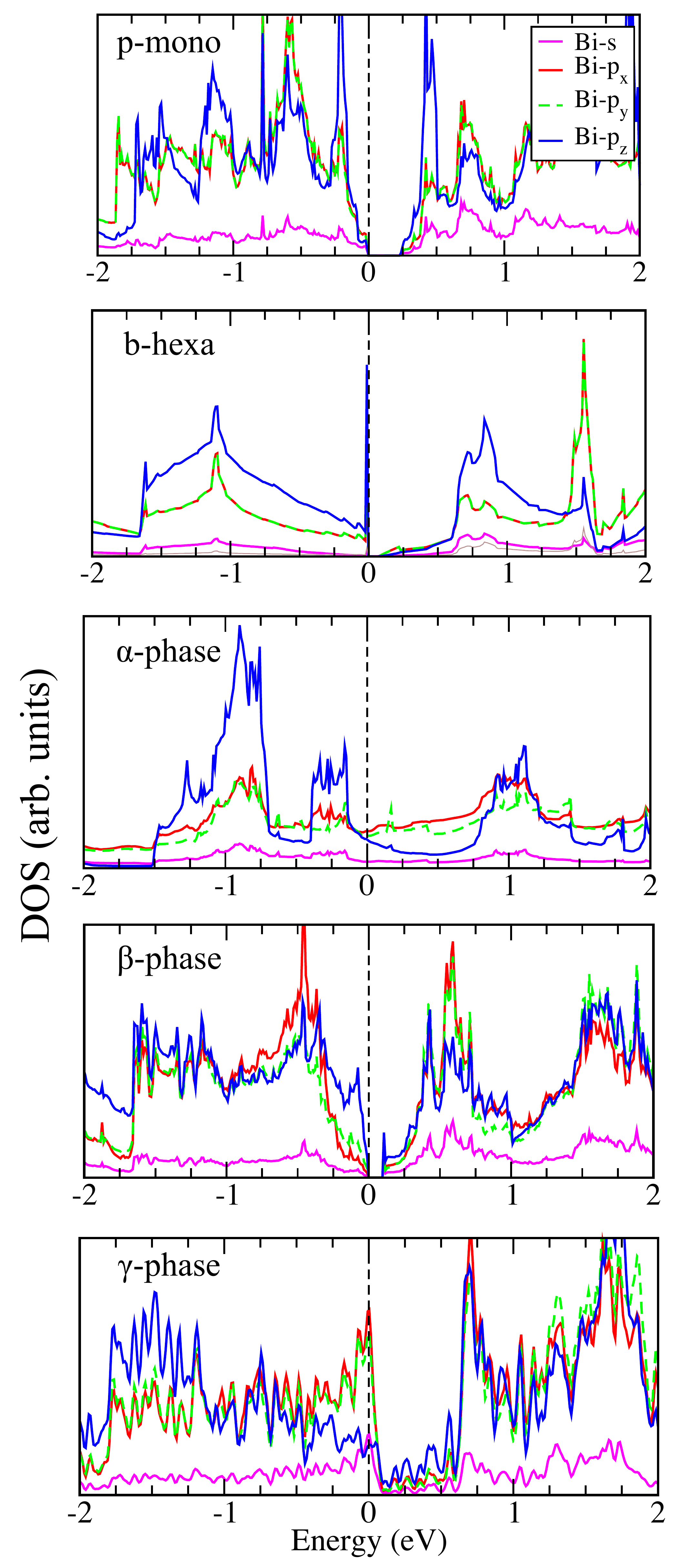}
 \caption{(Color online) Density of states (DOS) calculated with-SOC using a $k$-mesh of size $21\times21\times1$ for p-mono, b-hexa, $\alpha$-, $\beta$-, and $\gamma$-phases of bilayer Bi (arranged in a top-to-bottom order, respectively). }
 \label{fig:dos}
 \end{figure}

The orbital projected density of states (DOS), shown in Fig.~\ref{fig:dos}, suggests that the valence and conduction states are primarily composed of Bi $p_x$, $p_y$ and $p_z$ states, which is expected for Bi bilayers. DOS plots also confirm the semiconducting behavior of p-mono, b-hexa and $\beta$ phases, and metallic behavior of $\alpha$ and $\gamma$ phases. In b-hexa bilayer, we observe a sharp enhancement in the DOS of the occupied states at the Fermi-level, which is due to the Mexican-hat type shape of the highest valence band near the $\Gamma$ point. Such a divergence in the DOS can lead to an electronic instability, often resulting in structural distortions, magnetism, or superconductivity~\cite{SeixasPRL2016}. The presence of Dirac points yields a strong enhancement in the DOS of the $\gamma$-phase near the Fermi level. Since DFT predicts an underestimated electronic bandgap, we perform HSE06 calculations~\cite{HSE_2003} to obtain a better estimate. The HSE06 predicted bandgaps (without-SOC) for semiconducting p-mono, b-hexa and $\beta$ phases are 0.51\,eV,  0.72\,eV, and 1.16\,eV, respectively.

\section{Summary}
In summary, we report five low-energy crystal structures of bilayer Bi that are obtained from a systematic structural search in two-dimensions. In the lowest energy phase, Bi atoms prefer a puckered monoclinic structure instead of a hexagonal buckled bilayer structure. The energetic ordering from low to high is as follows: puckered monoclinic, buckled hexagonal, $\alpha$, $\beta$ and $\gamma$ phases. Except $\alpha$ and $\gamma$ phases, all other phases are dynamically stable. We find that the puckered monoclinic, buckled hexagonal, and $\beta$ phases are semiconducting, whereas the $\alpha$ and $\gamma$ phases exhibit metallic properties. We notice signatures of a van Hove singularity in the DOS of the buckled hexagonal bilayer near the Fermi-level. The $\gamma$ phase hosts topological type-I and type-II Dirac points together with the signatures of interesting topological Lifshitz transitions occurring near the Fermi-level. A structural distortion-mediated metal-insulator phase transition can be realized in the reported Bi bilayers.

\section{METHODS}
 \label{sec:methods}
 \subsection{Structure search algorithm}
The prediction of new crystal structures using advanced search methods has become a powerful tool for materials discovery and design. The computational prediction of new structures for a given atomic composition requires a systematic exploration of the multidimensional potential energy surface (PES), in order to find the global and local minima. In this work, we employ the minima hopping method (MHM)~\cite{Goedecker2004,Amsler2010} to carry out structural search calculations in a constrained configurational space for Bi. This method seeks local minima on the multidimensional PES, using an efficient dynamical algorithm, combining Density Functional Theory (DFT) to evaluate energy and forces, and short Molecular Dynamics (MD) simulations to help escape from a local minimum and explore new regions of the PES. The initial velocities during MD simulations are aligned along the soft mode direction to cross over low-energy barriers, thereby exploiting the Bell-Evans-Polanyi principle~\cite{roy_2009,Jensen}. The MHM employs a feedback mechanism to avoid revisiting local minima and to accelerate the search. More details of this method for structure search in 3D can be found in Refs.~\cite{amsler2014crystal, MHM_application-1, singh2016PCCP, Pavlic201715, singh2018thesis}. 

In order to perform structure searches in two dimensions, we add a confinement potential, to restrict the motion of atoms within a pre-defined thickness. Within this region, the confinement potential attains a zero value, but it increases quartically as we go beyond this pre-defined thickness (similar to the particle in a well problem). Thus, the search is restricted to find the low-energy arrangements of atoms in a constrained quasi-2D space. This approach has been successfully applied to identify two-dimensional forms of TiO$_2$~\cite{MaximilianCoM2017}. 

An extended version of the minima hopping method (MHM)~\cite{Goedecker2004, Amsler2010} was used to predict the low-energy structures of bilayer Bi. A two-dimensional confining potential $C(e,\textbf{r}_i^\alpha)$ was added to the target energy function to be optimized, where $\alpha$ denotes the axis $\alpha = \{x,y,z\}$ along the non-periodic direction, $\textbf{r}_i$ are the cartesian coordinates of the $N$ atoms in the system, and $e$ is the equilibrium positions along $\alpha$ at which the potential is centered. We used a sum of atomic contributions as the confinement function, which is zero within a cutoff region $r_c$ around $e$, while it has a polynomial form of order $n$ with amplitude $A$ beyond $r_c$: 

\begin{equation}
C^\alpha=\sum_{i=1}^N c(e,\textbf{r}_i^\alpha) \\
 \label{eq1}
 \end{equation}

where,
\begin{equation}
\begin{aligned}
c(e,\textbf{r}_i^\alpha) =
\begin{cases}
A(|e - \textbf{r}_i^\alpha|-r_c)^n, & \text{for} ~~ |e-\textbf{r}_i^\alpha|\geq r_c \\
0, & \text{otherwise}
\end{cases}
 \end{aligned}
 \label{eq2}
 \end{equation}
 
The derivatives with respect to the atomic coordinates $\textbf{f}_i=\frac{\partial C_i}{\partial \textbf{r}_i}$ and the cell vectors  $\sigma_i=\frac{\partial C}{\partial h_i}$ were taken fully into account during the local geometry optimizations, the MD escape trials, and for aligning the initial MD velocities along the soft mode directions, a process that we call softening. Thereby, the Bell-Evans-Polanyi principle is exploited to accelerate the search towards the low energy structures~\cite{roy_2009,Jensen}. Note that the above atomic positions are expressed in the reduced coordinates $\textbf{r}_i=h\textbf{s}_i$, and $h=(\textbf{a}, \textbf{b}, \textbf{c})$ is the matrix containing the lattice vectors. In our structural search runs we used a confinement potential centered along the lattice vector $\textbf{c}$ ($i.e.$ $\alpha=z$), with a cutoff $r_c=0.3$~\AA, $n=4$ and $A=0.1$~eV. The structures predicted from the MHM structural search runs were further re-optimized using a tighter convergence criteria of $k$-mesh sampling and energy cutoff for the plane wave basis set. 

\subsection{Ab-initio calculations}
Density Functional Theory (DFT) based first-principles calculations were carried out using Projector Augmented Wave (PAW) method as implemented in the {\sc VASP} software~\cite{Kresse1996, Kresse1999}. We considered fifteen valence electrons of Bi ($5d^{10}6s^{2}6p^{3}$) in the PAW pseudo-potential. The exchange-correlation energy was computed within the generalized gradient approximation using the PBE exchange-correlation functional as parametrized by Perdew-Burke-Ernzerhof~\cite{Perdew1996}. SOC was included self-consistently. We used 600\,eV as the kinetic energy cutoff of the plane wave basis set and a $11 \times 11 \times 1$ Monkhorst-Pack $k$-mesh was used to sample the reciprocal space for structural optimization. The electronic density of states (DOS) was calculated using a $k$-mesh of size $21 \times 21 \times 1$. We used a $\Gamma$-type sampling scheme for hexagonal structure and Monkhorst-Pack scheme was used to sample the Brillouin zone of all other structures. Structural relaxations were performed until all the atomic forces were less than $10^{-3}$\,eV/\AA, and $10^{-8}$\,eV was used as the energy convergence criterion for self-consistent DFT calculations. Phonon calculations (with-SOC) were performed using the finite-displacement approach, and the PHONOPY software~\cite{phonopy, Togo2008} was used to evaluate the force-constants. Depending upon the primitive unit cell, supercells of size $3 \times 3 \times 1$ or $4 \times 4 \times 1$ were used for phonon calculations. A vacuum of thickness larger than 14 \AA~was added to avoid any periodic interaction between two adjacent Bi bilayers.  The {\sc PYPROCAR} code was used to analyze the electronic band structures and spin-textures~\cite{PyProcar, sobhit2016PRB}.

\textit{Acknowledgments}: This work used the Extreme Science and Engineering Discovery Environment (XSEDE), which is supported by National Science Foundation grant number OCI-1053575. Additionally, the authors acknowledge support from Texas Advances Computer Center (TACC), Bridges supercomputer at Pittsburgh Supercomputer Center and Super Computing Systems (Spruce and Mountaineer) at West Virginia University (WVU). AHR and SS acknowledge support from National Science Foundation (NSF) DMREF-NSF 1434897, NSF OAC-1740111, and DOE DE-SC0016176 projects. SS acknowledges support from the Dr. Mohindar S. Seehra Research Award and the Distinguished Doctoral Scholarship at West Virginia University. ZZ acknowledges financial support by the Ramon y Cajal program (RYC-2016-19344), the Spanish MINECO (FIS2015-64886-C5-3-P), the CERCA programme of the Generalitat de Catalunya (grant 2017SGR1506), and by the Severo Ochoa programme (MINECO, SEV-2017-0706). MA acknowledges support from the Novartis Universit{\"a}t Basel Excellence Scholarship for Life Sciences and the Swiss National Science Foundation (Project No. P300P2-158407, P300P2-174475). MJV acknowledges funding by the Belgian FNRS (PDR G.A. T.1077.15-1/7), ULiege and the Communaut\'{e} Fran\c{c}aise de Belgique (ARC AIMED G.A. 15/19-09), and computational resources from the Consortium des Equipements de Calcul Intensif (CECI, FRS-FNRS G.A. 2.5020.11) and Zenobe/CENAERO funded by the Walloon Region under G.A. 1117545


\bibliography{Bi_mono_bibliography}

\end{document}